\documentclass[epj]{svjour}
\usepackage{amsmath,amssymb,amsfonts}
\usepackage{color}
\usepackage{epsfig}
\usepackage{graphicx}

\newcommand{\ptp}{Prog. Theor. Phys. }

\newcommand{\ijmpa}{Int. J. Mod. Phys. A }
\newcommand{\prl}{Phys. Rev. Lett. }
\newcommand{\ap}{Adv. Phys. }
\newcommand{\rmp}{Rev. Mod. Phys. }
\newcommand{\jmp}{J. Math. Phys. }
\newcommand{\pr}{Phys. Rev. }
\newcommand{\ssc}{Solid State Commun. }
\newcommand{\plb}{Phys. Lett. B }
\newcommand{\physrep}{Phys. Rep. }
\newcommand{\prb}{Phys. Rev. B }

\begin{document}

\title{
Fermionic renormalization group flow into phases with broken discrete 
% RG 28.10.05
% symmetry: charge-density wave model
symmetry: charge-density wave mean-field model
}
\author{Roland Gersch \inst{1} \and Carsten Honerkamp \inst{2} 
\and Daniel Rohe \inst{1}\and Walter Metzner\inst{1}}
\institute{Max-Planck-Institut f\"{u}r Festk\"{o}rperforschung, 
  Heisenbergstra\ss{}e 1, D-70569 Stuttgart, Germany \and
  Institut f\"{u}r Theoretische Physik,
  Universit\"{a}t W\"{u}rzburg,
  Am Hubland, D-97074 W\"{u}rzburg, Germany}
\authorrunning{Gersch, Honerkamp, Rohe, Metzner}
\titlerunning{RG flow into CDW state}
\abstract{
  We generalize 
  the application of the 
  functional renormalization group (fRG) method 
  for the fermionic 
  flow into the symmetry-broken phase to finite temperatures. 
  We apply the scheme to the case of a broken discrete symmetry: 
% RG 28.10.05:
%  the one-dimensional charge-density wave (CDW) model at half 
  the charge-density wave (CDW) mean-field model at half
  filling. We show how an arbitrarily small initial CDW 
  order parameter starts to grow at the CDW instability 
  and how it flows to the correct final 
  value, suppressing the divergence of the effective interaction
  in the fRG flow. 
  The effective
  interaction peaks at the instability and saturates at low energy
  scales or temperatures. The relation to the mean-field treatment, 
  differences compared to the flow for a broken continuous symmetry, 
  and the prospects of the new method are discussed.  
}
\date{Submitted 9 September 2005, published online 23 December 2005}
\maketitle
\section{Introduction}
\label{sec:intro}
Renormalization group (RG) techniques are a standard tool for the analysis of the low-energy physics of interacting fermions \cite{2001PThPh.105....1S,1997cond.mat..1012M,1994RvMP...66..129S}.
Although new effective degrees of freedom may be generated in the RG flow down to lower energies, perturbative RG flows with 
the initial degrees of freedom - in our case electrons - are considered the least biased approaches to the variety of 
competing and conspiring tendencies at low scales. 
Without appropriate self-energy corrections
these flows often diverge at a 
nonzero energy scale, and not all fermionic modes can be integrated out. 
These divergences are signatures of potential phase 
transitions, typically involving some kind of symmetry breaking. 
Hence, these flows still contain useful physical information. 
In many cases the physics below the critical scale can be explored by other
means, e.g. by a 
mean-field treatment \cite{reissrohemetzner}, 
by a bosonic description \cite{2004PhRvB..70l5111B}
or by exact diagonalization of a restricted low-energy
Hamiltonian \cite{2004PhRvL..92c7006L}.
Yet, the 
latter methods always require some sort of simplification of the low-energy action, and typically the final 
results depend on the scale where the transition from one treatment to the other is performed. Therefore,  
it is desirable to have an RG scheme which can be continued beyond the noted divergence and which allows a 
controlled access to the low-energy phase. 

Recently, Salmhofer et al. \cite{2004PThPh.112..943S} have proposed an extension of the fermionic functional RG flow into 
the symmetry-broken regime. The key idea is to include an arbitrarily small
symmetry breaking field, which we include in the RG as a symmetry-breaking component in the initial condition 
for the fermionic self-energy. This component grows rapidly at the critical scale and prevents a true divergence of the interactions.
Therefore, the flow can be continued down to the lowest
scales.
Spontaneous symmetry breaking is captured by sending the
external symmetry-breaking field to zero after integrating
the flow.
For mean-field models like the reduced BCS model, this
procedure yields the exact solution already within a
suitably chosen one-loop truncation of the vertex flow 
\cite{2004PThPh.112..943S}. 

Of course this  method should now be extended to other models and other
physical situations. In particular, the hope is that non-mean-field models 
like the two-di\-men\-sional Hubbard model will be treatable. 
There, the method would yield the flow into sym\-me\-try-broken phases corresponding to correlations 
for which the initial interaction does not contain an attractive component and where the attraction is generated during the flow.
For example, this is the case for $d$-wave pairing in the Hubbard model on the
two-dimensional square lattice
near half filling. Furthermore, the non-mean-field parts 
of the interactions would not need to be dropped and their influence could be studied.

Before proceeding to such more complicated problems, we extend the
application of the formalism to nonzero temperatures in
the context of the charge-density wave (CDW) mean-field model on a $d$-dimensional hypercubic lattice at half band filling.
Again, we closely monitor the correspondence to mean-field theory. 
This example is interesting because, in contrast to the BCS model studied in Ref. \cite{2004PThPh.112..943S}, 
only a discrete symmetry is broken. Hence we learn about the differences between the RG flows for broken continuous and discrete symmetries.

This paper is organized as follows. In Sec. \ref{sec:model} 
we introduce the CDW model which we analyze in the remainder. 
In Sec. \ref{sec:exact}
we present an exact solution in the thermodynamic limit employing a resummation
of perturbation theory. 
We briefly introduce the renormalization group technique
in Sec. \ref{sec:RGsetup} and write down the functional RG equations
for the self-energy and the effective interaction. 
In Sec. \ref{sec:RGT0}, these equations are discussed in the case of vanishing 
temperature, where they can be presented in a considerably simpler way.
The finite-temperature equations are treated numerically in Sec.
\ref{sec:RGfT}. Conclusions and outlook are presented in 
Sec. \ref{sec:conclusion}.

\section{Model}
\label{sec:model}
We consider spinless fermions on a $d$-dimensional hyper-cubic lattice with $N$
sites labeled by $\vec{x}$. The kinetic energy shall be given by a 
tight-binding dispersion with
nearest-neighbor hopping amplitude $t$. 
Further we assume a repulsive nearest-neighbor density-density
interaction $V_0$. Then, the Hamiltonian reads
\begin{equation}
  \label{eq:ham_dir}
  {H}
  =
  -t\sum_{\vec{x},\vec{n}} ( c_{\vec{x}}^\dagger c^{}_{\vec{x}+\vec{n}} + h.c. )
  +
  V_0 \sum_{\vec{x},\vec{n}}
    c_{\vec{x}}^\dagger c^{\phantom\dagger}_{\vec{x}}  c_{\vec{x}+\vec{n}}^\dagger c^{\phantom \dagger}_{\vec{x}+\vec{n}}.
\end{equation}
The sum over $\vec{n}$ runs over the $d$ unit vectors of the $d$-dimensional hypercubic lattice with lattice constant set to one. 
Now we assume half filling with an average of one particle per two lattice
sites, $\langle n_{\vec{x}} \rangle = 1/2 $.  In this case it is easy to see that the 
nearest-neighbor density-density repulsion will favor a  charge
density wave (CDW), i.e. a periodic arrangement of the
particles in which the probability of two particles being nearest neighbors is
reduced. This tendency competes with the hopping term of the kinetic energy and with the entropy, which are minimized and maximized, respectively, for an equal population of all sites. In the case where a CDW is formed with a fixed modulation amplitude $n_{\mathrm{CDW}}$, the charge density
$\langle n_{\vec{x}} \rangle $ takes only two
values, $1/2 \pm n_{\mathrm{CDW}}$, on the two inter-penetrating sublattices with doubled unit cells. 
Therefore, the half-filled CDW state breaks 
a discrete Ising-type symmetry, and the degeneracy of the ordered state is twofold. 
While in one dimension such order can only occur in the ground state, in two
and higher dimension it is possible to have a nonzero critical temperature $T_c$. 

For general band filling, a CDW with an incommensurate modulation wave-vector generates infinitely many different values of average populations on the lattice sites. This corresponds to an arbitrary phase of the density modulation at an arbitrary reference lattice site. Hence in the incommensurate case the CDW breaks 
a continuous symmetry. %U(1)-symmetry.  

Continuing with the half-filled case, we Fourier-trans\-form the Hamiltonian using
\begin{equation}
  \label{eq:fourier_conventions} 
  c_{\vec{x}} 
  = 
  \frac{1}{\sqrt{N}} \sum_{\vec{k}}
  e^{i\vec{k}\vec{x}}c^{}_{\vec{k}}
\end{equation}
and obtain
\begin{align}
   \nonumber 
   {H}
   =&
   - 2t\sum_{\vec{k}}
   L(\vec{k}) \, c_{\vec{k}}^\dagger c^{\phantom{\dagger}}_{\vec{k}}
   \\ &+
   \frac{V_0}{N}\sum_{\vec{k}_1,\vec{k}_2,\vec{q}}
  L(\vec{q}) \,
   c_{\vec{k}_1}^\dagger c^{\phantom\dagger}_{\vec{k}_1-\vec{q}} c_{\vec{k}_2}^\dagger c^{\phantom\dagger}_{\vec{k}_2+\vec{q}}.
   \label{eq:ham_full}
\end{align}     
Here we have introduced $L(\vec{k})=\sum_{i=1}^d \cos ( k_i )$. 
Abbreviating the dispersion as $\xi (\vec{\vec{k}})= -2t \sum_{i=1}^d \cos (k_i) = -2t L(\vec{k})$, 
we note that the nesting condition $\xi (\vec{\vec{k}}) = - \xi (\vec{\vec{k}}+ \vec{Q})$ is fulfilled for
the wave-vector $\vec{Q}=(\pi,\dots,\pi)$. This causes a divergence of the
non-interacting charge response at the nesting wave-vector for $T\to 0$. 
If the interactions are treated 
in the random phase approximation, the charge response
at $\vec{Q}$
will actually diverge at a finite temperature, giving an estimate of $T_c$
for the charge-density wave formation. In a mean-field treatment of the
symmetry-breaking in which the interaction term is decoupled 
with an alternating charge density, anomalous particle-hole pairing 
expectation values
$\langle c^\dagger_{\vec{\vec{k}}+\vec{Q}} c^{\phantom\dagger}_{\vec{\vec{k}}} \rangle$
will become nonzero below the transition temperature. 

Although this physical picture of a CDW transition is essentially correct
regarding the ground state properties, the model \eqref{eq:ham_full} cannot be
solved exactly. 
This changes if we reduce the interactions by only keeping processes
that change both particle's momenta by $\vec{Q}$,
\begin{align}
  \nonumber 
  {H}_\mathrm{red}
  =&
  \sum_{\vec{k}} \xi (\vec{k})
   \, c_{\vec{k}}^\dagger c^{\phantom\dagger}_{\vec{k}}
  \\& -
  \frac{V_0}{N}\sum_{\vec{k}_1,\vec{k}_2}
  c_{\vec{k}_1}^\dagger c^{\phantom\dagger}_{\vec{k}_1-\vec{Q}} c_{\vec{k}_2}^\dagger c^{\phantom\dagger}_{\vec{k}_2+\vec{Q}}.
  \label{eq:ham_reduced}
\end{align}
Now the  interaction between fermions on the lattice sites 
$\vec{x}$ and $\vec{x}'$ corresponds to an infinite range 
density-density interaction with oscillating sign,
% RG 28.10.05
% $-V_0 \cos [\vec{Q} (\vec{x}-\vec{x}')]  n_{\vec{x}} n_{\vec{x}^\prime}$.
\\$-V_0N^{-1} \cos [\vec{Q} (\vec{x}-\vec{x}')]  n_{\vec{x}} n_{\vec{x}^\prime}$.

Analogous to the reduced BCS pairing model \cite{muehlschlegel}, in the reduced model  \eqref{eq:ham_reduced} mean-field theory becomes exact in the thermodynamic limit $N\to \infty$ and there is a CDW ordered state below a critical $T_c>0$ in any dimension $d$. 
For the half-filled commensurate case, the CDW state is twofold degenerate and the electronic order parameter $n_{\mathrm{CDW}}$ is real. 

Next we include a real external field $\Delta_{\rm ext} \cos (\vec{Q}\vec{x})$. This term breaks the translational symmetry of the original Hamiltonian explicitly and lifts the degeneracy of the two CDW ground states. In momentum space this term couples to pairs of particles whose momenta differ by 
$\vec{Q}=(\pi, \dots, \pi)$. 
 If $\Delta_{\rm ext}$ is very small
compared to all other relevant energy scales,
it does not change macroscopic observables away from the critical
temperature. However, it allows us to integrate the
RG differential equations over all scales without
encountering divergences.
The Hamiltonian including the external field reads
\begin{align}
  H_{\rm red}=&H_{\rm kin}+
  \sum_{\vec{k}} %_{-N/2+1}^{N/2}
  \Delta_{\rm ext}c^\dagger_{\vec{k}} c^{\phantom\dagger}_{\vec{k}+\vec{Q}}
  \nonumber\\
  &-
  \frac{V_0}{N}\sum_{\vec{k}_1,\vec{k}_2}%=-N/2+1}^{N/2}
  c^\dagger_{\vec{k}_1} c^{\phantom\dagger}_{\vec{k}_1+\vec{Q}} 
  c^\dagger_{\vec{k}_2} c^{\phantom\dagger}_{\vec{k}_2-\vec{Q}}
  .
  \label{eq:ham_red}
\end{align}

We introduce a frequency-space field-integral representation in the usual way.
Writing $T$ for temperature, we
define the fermionic Matsubara frequencies $\omega_j:=(2j+1)\pi T$ and
obtain Grassmann fields $\psi_{\vec{k},\omega_n},\bar{\psi}_{\vec{k},\omega_n}$.
Introducing the Nambu-like notation
\begin{eqnarray*}\Psi_{\vec{k},\omega_n}
&=
\left(
  \begin{matrix}
    \psi_{\vec{k},\omega_n}\\
    \psi_{\vec{k}+\vec{Q},\omega_n}
  \end{matrix}
\right),
\,\,\\
\bar{\Psi}_{\vec{k}+\vec{Q},\omega_n}
&=
\left(
  \begin{matrix}
    \bar{\psi}_{\vec{k},\omega_n}& 
    \bar{\psi}_{\vec{k}+\vec{Q},\omega_n}
  \end{matrix}
\right) \, , 
\end{eqnarray*}
the partition function reads
\begin{equation*}
  \nonumber 
  {\cal Z} 
  = 
  \int{\cal D}(\bar{\psi},\psi)\exp
  \left[
    \frac{1}{2}
    T \sum_{n,\vec{k}}
    \bar{\Psi}_{\vec{k},\omega_n}
    {\mathbb Q}(\xi_{\vec{k}},\omega_n)
    \Psi_{\vec{k},\omega_n}\right.
\end{equation*}
\begin{equation}
  +V_0
    \frac{T^3}{N}\sum_{\substack{n_1,n_2,n_3\\\vec{k}_1,\vec{k}_2}}
    \bar{\psi}_{\vec{k}_1,\omega_{n_1}}
    \psi_{\vec{k}_1+\vec{Q},\omega_{n_3}}\left.
    \vphantom{\sum_{n\in{\mathbb Z}}\int_{-\pi}^{\pi}}
    \bar{\psi}_{\vec{k}_2,\omega_{n_2}}
    \psi_{\vec{k}_2-\vec{Q},\omega_4}
  \,\right] 
  \label{eq:part_sum}.
\end{equation}
Here, $\omega_4=\omega_{n_1+n_2-n_3}$ and 
${\mathbb Q}$ is the matrix inverse of the free Green's function
(see e.g. \cite{LeeRiceAnderson1974})
\begin{equation}
  \label{eq:green_free}
  {\mathbb G}_0(\xi,\omega_n)
  =
  \frac
  {1}
  {\omega_n^2+\xi^2+\Delta_{\rm ext}^2}
  \left(
    \begin{matrix}
      -i\omega_n-\xi & -\Delta_{\rm ext} \\
      -\Delta_{\rm ext} & -i\omega_n+\xi
    \end{matrix}
  \right).
\end{equation}
We have dropped the momentum argument of $\xi$ for brevity.
The minus sign in front of the first $\xi$ arises from
the $\vec{Q}$-anti-periodicity of the cosine. The doubling of the number of degrees of freedom by introducing the Nambu notation is compensated by the $1/2$ in front of the quadratic part of the action. Alternatively,   
 we could integrate over only half the
Brillouin zone. Due to the explicit symmetry breaking, there is a nonzero
off-diagonal component in the propagator. If we apply mean-field theory by
decoupling the interaction term in the CDW
channel, the CDW order parameter 
due to the interaction, $\Delta_{\mathrm{ia}}$,
will add to $\Delta_{\mathrm{ext}}$. 

\section{Mean-Field Solution and Resummation}
\label{sec:exact}
In this section, we resum the perturbation expansions for the
self-energy and the effective interaction 
in the thermodynamic 
limit $N\to\infty$. For our model this is equivalent to mean-field theory in the CDW amplitude with modulation wavevector $\vec{Q}$.  

Due to the explicit symmetry breaking with $\Delta_{\rm ext} \not= 0$  there is a nonzero frequency independent off-diagonal self-energy which scatters a fermion with wavevector $\vec{k}$ to $\vec{k+Q}$ and vice versa. 
Therefore, besides normal (diagonal) wave-vector conserving propagators, also
anomalous (off-diagonal) propagators appear in the diagrams of the
perturbation theory. As indicated above, compared to the non-interacting propagator in \eqref{eq:green_free}, the effect of the interactions is that we have to supplement $\Delta_{\rm ext}$ with an off-diagonal self-energy 
$\Delta_{\rm ia}$ caused by the interaction. 
In the following we determine its value. 

\begin{figure}[htbp]
  \centering
  \input{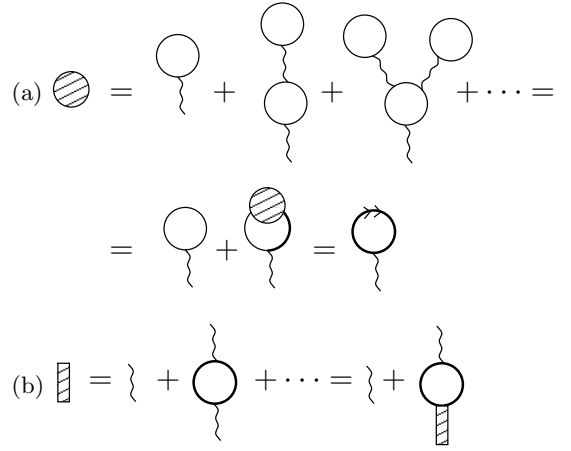}
  \caption{Resummation of the perturbation expansion of
    the reduced charge-density-wave model: 
    (a) self-energy (hatched circle), 
    and (b) effective interaction (hatched rectangle). 
    Undirected internal lines
    carry a summation over their Nambu-like indices in addition
    to the frequency summation and momentum integration. 
    Double-arrowed lines correspond to off-diagonal propagator elements,
    bold lines to full and thin lines to bare
    propagators.}
  \label{fig:resum}
\end{figure}

Since the interaction term in \eqref{eq:ham_red} includes
a prefactor $1/N$, only diagrams having one summation over
all degrees of freedom per interaction line contribute in
the thermodynamic limit. 
This together with the special structure of the interaction 
entails that only the bubble diagrams shown in Fig. 1(b) contribute to the perturbation series for the effective interaction  in second order of the bare interaction. In all other second order diagrams the restricted interaction implies that the internal momentum is fixed by an external momentum. Then the contribution of these diagrams is $O(1/N)$ and vanishes for $N\to \infty$. 
Furthermore, 
since each bubble in the remaining diagrams transfers a momentum
of $\vec{Q}$ due to the special structure of the bare interaction,
only effective interaction processes with a momentum
transfer of $\vec{Q}$ exist.
This causes the normal diagonal self-energy to vanish as well. 
These arguments can be iterated to arbitrary order, and hence only the bubble chains indicated in Fig. 1(b) contribute to the perturbation expansion of the effective interaction. In particular, no anomalous effective interactions which violate the original translation symmetry are generated. This is different away from half-filling where pairs of internal anomalous propagators 
can generate effective interactions which violate momentum conservation
by adding even multiples of the modulation wavevector $\vec{Q}$ to their incoming total momentum. In the half-filled case $2\vec{Q}=(2\pi, \dots, 2\pi)$ is a reciprocal lattice vector, and therefore no new effective interactions appear.  

In Fig. \ref{fig:resum}(a) we show the perturbation expansion for the off-diagonal self-energy. Again, due to the special structure of the interaction
 only those diagrams which retain one momentum integration per interaction line contribute to the perturbation expansion of the self-energy in the thermodynamic limit. 
Furthermore, the interaction is nonzero only for momentum transfer $\vec{Q}$. 
Thus, in the thermodynamic limit only 
diagrams constructed from tadpole diagrams contribute. 
If a tadpole diagram is connected to at most one other tadpole diagram, only 
the terms off-diagonal in the Nambu-like space are non-vanishing. 
This is because the outgoing momentum of such an off-diagonal line 
equals the incoming momentum plus $\vec{Q}$.
The exact resummation is equivalent to 
self-consistent Hartree-Fock theory of which the Fock contributions
vanish in the thermodynamic limit. Another way of saying this is that Har\-tree mean-field theory is exact in this model.

In the following, we assume the density of states to
be constant over the Brillouin zone and equal to its value at
the Fermi energy to remove dimension-specific effects.
Evaluating Fig. \ref{fig:resum}(a) and denoting
the self-energy due to the interaction by $\Delta_{\rm ia}$, 
the band edge by $W$, the density of states at the Fermi energy $\rho_0$,
$\Sigma:=\Delta_{\rm ia}+\Delta_{\rm ext}$ as well as
$E:=\sqrt{\xi^2+\Sigma^2}$,
we obtain the gap equation (compare e.g. 
\cite{RiceStrassler1973})
\begin{equation}
  \label{eq:gap_full}
  \Sigma-\Delta_{\rm ext}
  =
  V_0\Sigma\rho_0
  \int_{0}^W{\rm d}\xi
  \frac
  {\tanh(E/2T)}
  {E}
\end{equation}
which resembles the BCS \cite{1957PhRv..108.1175B}
and excitonic insulator \cite{1967PhRv..158..462J} 
gap equations.
If we set $T=\Delta_{\rm ext}=0$, 
solving \eqref{eq:gap_full} analytically yields (compare e.g. 
\cite{1957PhRv..108.1175B,1985PhR...119..117G})
\begin{equation}
  \label{eq:gap_simp}
  \Sigma
  =
  2t
  \frac{1}{\sinh
  \left(
    1/V_0\rho_0
  \right)}
  \overset{V_0\rho_0 \ll 1}{\approx}
  4t\exp
  \left(-
    \frac{1}{V_0\rho_0}
  \right).
\end{equation}
Similar equations are found in the BCS theory for superconductors 
\cite{1957PhRv..108.1175B}.
\begin{figure}[htbp]
  \centering
  \includegraphics{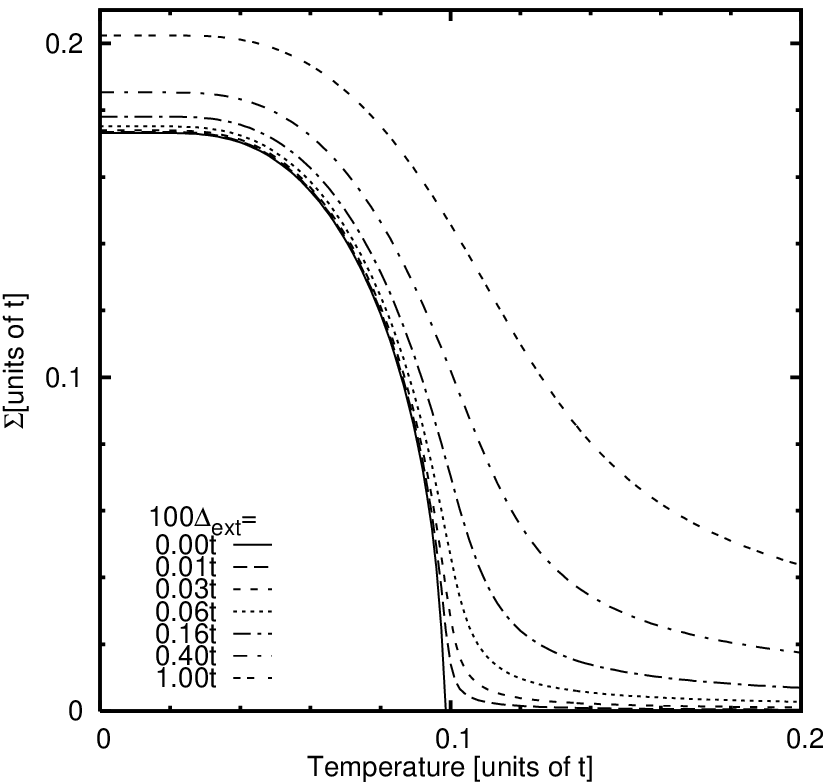}
  \includegraphics{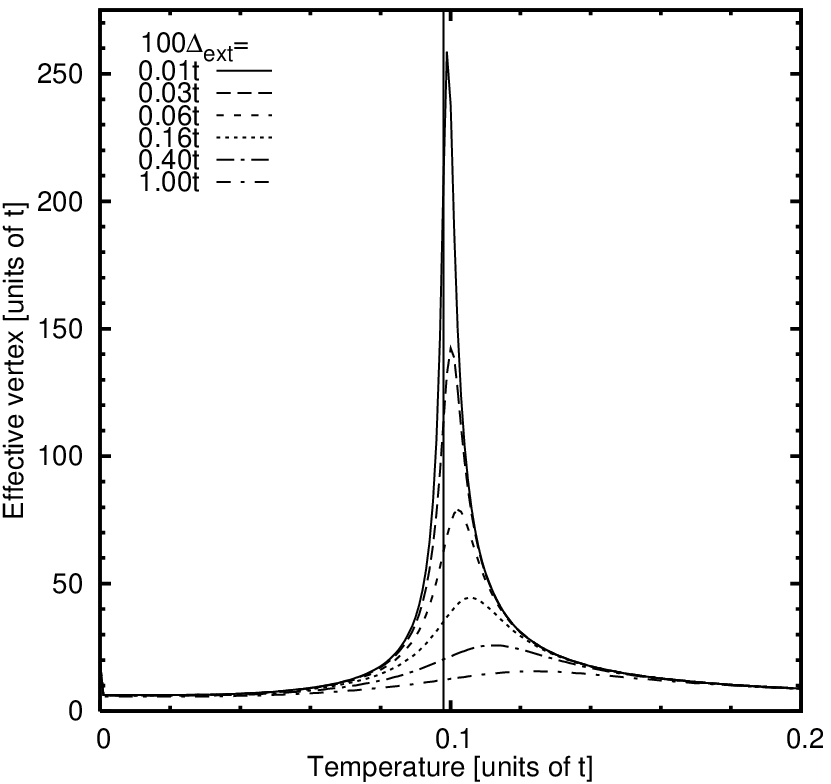}
  \caption{(top) Solutions $\Sigma(T)$ of the gap 
    equation for $V_0=2t$ and 
    small to intermediate  
    initial self-energies $\Delta_{\rm ext}$. 
    (bottom) Temperature dependence of the effective
    interactions calculated by resumming the perturbation
    expansion for $V_0=2t$
    and small to intermediate initial self-energy. 
    $\rho_0=1/2\pi$
    for all numerical calculations.
  }
  \label{fig:VRPA_T_U_0}
\end{figure}

We now turn to the effective interaction with zero frequency transfer. As argued above, only fermionic bubbles contribute and all other diagrams vanish in the thermodynamic limit $N\to \infty$. In the notation, we omit the arguments of
the hyperbolic functions originating from the Fermi distribution from
now on:
they are always $E/2T$.
Abbreviating the bubble integral
\begin{equation}
  B
  :=
  \rho_0\int_0^W
  \frac%
  {{\rm d}\xi}%
  {E^2}
  \left[
    \vphantom{
      \frac{\Sigma^2}{2T}
      \cosh^{-2}
      +
      \frac{\xi^2}{E}
      \tanh
    }
  \right.
  \frac{\Sigma^2}{2T}
  \cosh^{-2}
  +
  \left.
    \vphantom{
      \frac{\Sigma^2}{2T}
      \cosh^{-2}
      \left(
        \frac{E}{2T}
      \right)
      +
      \frac{\xi^2}{E}
      \tanh
      \left(
        \frac{E}{2T}
      \right)
    }
    \frac{\xi^2}{E}
    \tanh
  \right],
  \label{eq:bubble}
\end{equation}
the evaluation of Fig. \ref{fig:resum}(b) yields
\begin{equation}
  \label{eq:vertex}
  V
  =
  \frac
  {V_0}
  {1-V_0B}.
\end{equation}
If $\Delta_{\rm ext}=0$, we can define the
critical temperature $T_c$ via the condition that
$1=V_0B(T_c)$, i.e. that the denominator
of \eqref{eq:vertex} vanishes and $V$ diverges. By analyzing \eqref{eq:bubble} 
we note that this criterion for $T_c$ coincides 
with the temperature below which we can find a nonzero 
solution of the gap equation \eqref{eq:gap_full}.
For small $V_0$ and $\Sigma=0$ we find 
$1.76T_c\approx 4t\exp(-1/V_0\rho_0)\approx \Sigma(T=0)$ 
(see e.g. \cite{1985PhR...119..117G}). This is
true when the approximation in \eqref{eq:gap_simp} is valid.

As a finite $\Delta_{\rm ext}$ will later allow us to 
integrate the renormalization group equations over all
scales, we are interested in the dependence of the
mean-field solutions \eqref{eq:gap_full} and \eqref{eq:vertex}
on $\Delta_{\rm ext}$. In particular, we will illustrate
where the system including a finite symmetry-breaking field 
\eqref{eq:ham_red} is practically indistinguishable from the system 
without symmetry-breaking field, \eqref{eq:ham_reduced}.
To this end, we show in the upper part of Fig. \ref{fig:VRPA_T_U_0} 
convergence
of $\Sigma^{\Delta_{\rm ext}}(T)$ to $\Sigma(T)$ for
$\Delta_{\rm ext}\to 0$. 
The convergence is worst in the vicinity of a kink that
can be discerned at $T\approx 0.1t$ in the graph for $\Delta_{\rm ext}=0$.
We would like to be more concrete about the temperature range 
in which the system including a symmetry-breaking field can be used
to approximate the system without symmetry-breaking field.
To determine more precisely the temperature range in which the
symmetry-breaking field has little impact on the physics of the system,
we consider the dependence of the effective 
interaction on both temperature and the symmetry-breaking field as
plotted in the lower part of Fig. \ref{fig:VRPA_T_U_0}. 
Its singularity is regularized by 
a nonzero $\Delta_{\rm ext}$ which cuts off the integrand of
the bubble integral \eqref{eq:bubble}. 
The figure also shows a strong suppression
of the effective interaction peak with increasing $\Delta_{\rm ext}$ (the dependence of the
peak height on $\Delta_{\rm ext}$ is plotted in Fig. 
\ref{fig:sigma_deviation}).
Its location converges to $T_c$ from above for $\Delta_{\rm ext}\to 0$.
In the figure, we also spot convergence of $V^{\Delta_{\rm ext}}(T)$
to $V(T)$ for $T\ll T_c$ and $T\gg T_c$. We see that 
$V^{\Delta_{\rm ext}}(T)\approx V(T)$ if $T$ is outside the double width
at half maximum of $V^{\Delta_{\rm ext}}(T)$.
If $T$ is outside said width and also below $T_c$, we learn
from the upper part of the figure that
$\Sigma^{\Delta_{\rm ext}}(T)\approx\Sigma(T)$.
This is thus the temperature region in which the initial
symmetry-breaking field does not appreciably change the physics.

\section{Renormalization Group Setup}
\label{sec:RGsetup}
We will now apply the one-particle irreducible (1PI) 
functional renormalization group scheme to the charge-density-wave.
The scheme is described in \cite{2001PThPh.105....1S} (see
also \cite{1993PhLB..301...90W}) and
is here employed in the version 
suggested by Katanin \cite{2004PhRvB..70k5109K}.
In this version, the differential equations for the self-energy
and four-point function constitute a closed system. 
The special structure of our bare interaction will allow us to further
simplify the equations. We will see that in contrast to the BCS flow
\cite{2004PThPh.112..943S}, no anomalous effective interactions 
are generated. In later sections, we will verify numerically
that the fRG reproduces the resummation results.

Following \cite{2001PThPh.105....1S}, we introduce a positive
real parameter $s$ and a cutoff $\Lambda:=
2t\exp(-s)$. 
In our calculations, all modes satisfying $\left|\xi\right|>\Lambda$ 
have been integrated out. 
The effective interaction and self-energy at this scale
can thus be interpreted as parameters of an effective theory for
a reduced system with smaller bandwidth.
We start the flow at $s=0$ $(\Lambda=2t)$, where
all modes have yet to be integrated out, and integrate the flow down 
to $\Lambda=0$.
To analytically implement
this procedure, we introduce $\chi(\xi,\Lambda)$ as a placeholder
for any cutoff function,
 which shall have range $[0,1]$, assume the value $1/2$ when $\Lambda=\xi$, 
approach $0$ when reducing $\xi$ below $\Lambda$ and
$1$ when increasing $\xi$ above $\Lambda$.
We replace $\mathbb Q(\xi)$ 
by ${\mathbb Q}(\xi)/\chi(\xi,\Lambda)$, suppressing low-energy modes
in the field integral and rendering the self-energy and effective interaction
scale-dependent. In the following, dots over symbols 
will denote differentiation
with respect to $s$. In the diagrammatic representations,
a heavy line represents a scale-dependent full propagator 
(calculated with the scale-dependent self-energy),
a hatched circle represents the scale-dependent self-energy
and a hatched rectangle represents the scale-dependent effective interaction.
We furthermore introduce the single-scale propagator
${\mathbb S}:={\mathbb G}\dot{\mathbb Q}{\mathbb G}$ (all scale-dependent).
A line drawn through propagator lines indicates an $s$-derivative.

\begin{figure}[htbp]
  \centering
  \input{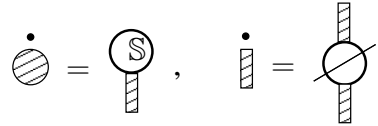}
  \caption{The hierarchy of flow equations for the self-energy 
    and the effective interaction.}
  \label{fig:RG}
\end{figure}

In the flow of the effective interaction, having  $\partial_s({\mathbb G}{\mathbb G})$ \cite{2004PhRvB..70k5109K} instead of  ${{\mathbb S}{\mathbb G}}+{{\mathbb G}{\mathbb S}}$ as in the original 1PI RG scheme \cite{2001PThPh.105....1S} is crucial in order to be able to follow the flow down to $\Lambda=0$.
In the Katanin version of the 1PI scheme, the exact hierarchy
of RG differential equations with the exception of the first
equation is written using only
full four-point functions, full propagators and scale\--\-differ\-entia\-ted
full propagators on the right-hand sides. 
The differential equations for the self-energy
and the four-point function thus constitute a closed system.
In particular, the contribution to the flow
of the four-point function which involves the six-point
function in the standard 1PI scheme \cite{2001PThPh.105....1S}
is taken into account in the Katanin scheme
by diagrams involving only four-point functions and full
or scale-differentiated full propagators, respectively.
The part with overlapping loops of this contribution 
vanishes for the special 
structure of the interaction and the thermodynamic limit considered 
here. 
The remaining part of this contribution is taken into
account by the replacement of the single-scale propagator
by the scale-differentiated full propagator.
Hence, the graphical form of the fRG differential equations is as shown in
Fig. \ref{fig:RG}. 

In the flow of the effective interaction (right part of Fig. \ref{fig:RG}), 
the summation over the
`Nambu' indices of the internal lines  includes normal and anomalous
propagators. Analogous to the perturbation expansion in the previous section,
%RG 28.10.05
%in the limit $N\to \infty$ the special form of the initial interaction is
in the limit $N\to \infty$ the special momentum structure of the initial 
interaction is
conserved by the RG flow of the effective interaction. In the present half-filled case, no
new effective interactions with different external legs are generated. 
This is in contrast
to the BCS pairing model, where U(1)-symmetry-breaking vertices
with four incoming or four outgoing legs are generated.
Linear combinations of normal and anomalous components of the effective interacion could be identified with `amplitude' and `phase' vertices.  In the half-filled CDW model the effective interaction behaves like the `amplitude' vertex of the BCS problem, while the `phase' vertex is missing as there is no breaking of a continuous symmetry.    
 The self-energy only contains the off-diagonal part of the single-scale propagator, again due to the restricted form of the interaction.
For a derivation see \cite{2001PThPh.105....1S} and 
\cite{2004PThPh.112..943S}. Ref. \cite{2004PThPh.112..943S} also provides
an analytical proof of the equivalence of the flow equations 
(Fig. \ref{fig:RG}) and the resummation equations (Fig. \ref{fig:resum}).

Evaluating the diagrams of Fig. \ref{fig:RG}, we 
suppress scale and
frequency dependence in the notation and write
$E:=\sqrt{\xi^2+\chi^2\Sigma^2}$. The RG differential
equations now read
\begin{align}
  \dot{V}=&
  \rho_0 V^2\left\{
    \frac{\Lambda}{E}
    \left.\left(
        \frac{\Lambda^2}{E^2}\tanh
        +
        \frac{\Sigma^2\beta}{2E}\cosh^{-2}
    \right)\right|_{\substack{\xi=\Lambda\\\chi=1}}-
  \right.
  \nonumber
  \\&
  \quad\left.
    \int_\Lambda^W\frac{{\rm d}\xi}{E}
    \frac{\Sigma\dot{\Sigma}}{E^2}
    \left[
      \frac{3\xi^2}{E^2}
      \left(
        \tanh
        -
        \frac{\beta E}{2}\cosh^{-2}
      \right)
    \right.
  \right.
  \nonumber
  \\&
  \qquad\left.
    \left.
      +\beta^2\Sigma^2\frac{\tanh}{2\cosh^2}
    \right] 
    \vphantom{    \left.\left(
          \frac{\Lambda^2}{E^2}\tanh
          +
          \frac{\Sigma^2\beta}{2E}\cosh^{-2}
        \right)\right|_{\substack{\xi=\Lambda\\\chi=1}}-
    }
  \right\},
  \label{eq:RG_vertex_full}
\end{align}
\begin{align}
  \dot{\Sigma}
  =
  V\left.\frac{\Sigma\rho_0\Lambda}{E}\tanh\right|_
  {\substack{\xi=\Lambda\\\chi=1}},
  \label{eq:RG_sigma_full}
\end{align}
where we have used 
${\mathbb S}=\dot{\chi}\partial{\mathbb G}/{\partial \chi}$,
the limit of a sharp cutoff, and (3.19) of 
\cite{morris94} 
(The simplification of the equations obtained using
the limit of a sharp cutoff is due to Tilman Enss).
The initial conditions are 
$\Sigma\left|_{s=0}\right.=\Delta_{\rm ext}$ and $V\left|_{s=0}\right.=V_0$.
If we choose $\Delta_{\rm ext}=0$, $\Sigma$ remains zero for all scales. 
Then only the first term in \eqref{eq:RG_vertex_full} remains, 
and the flow will diverge at a nonzero 
scale $\Lambda_c$.
We would thus be neither able to integrate over all scales nor compare our
results to the resummation of section \ref{sec:exact}. We therefore
always study the RG equations for a finite $\Delta_{\rm ext}$,
usually $10^{-4}t$. 
We will show that we can thereby circumvent the divergence of the effective interaction and still get arbitrarily close to the exact mean-field results without explicit symmetry breaking in the temperature range determined in
section \ref{sec:exact}. 

\section{Renormalization Group at Zero Temperature}
\label{sec:RGT0}
\begin{figure}[htbp]
  \centering
  \includegraphics{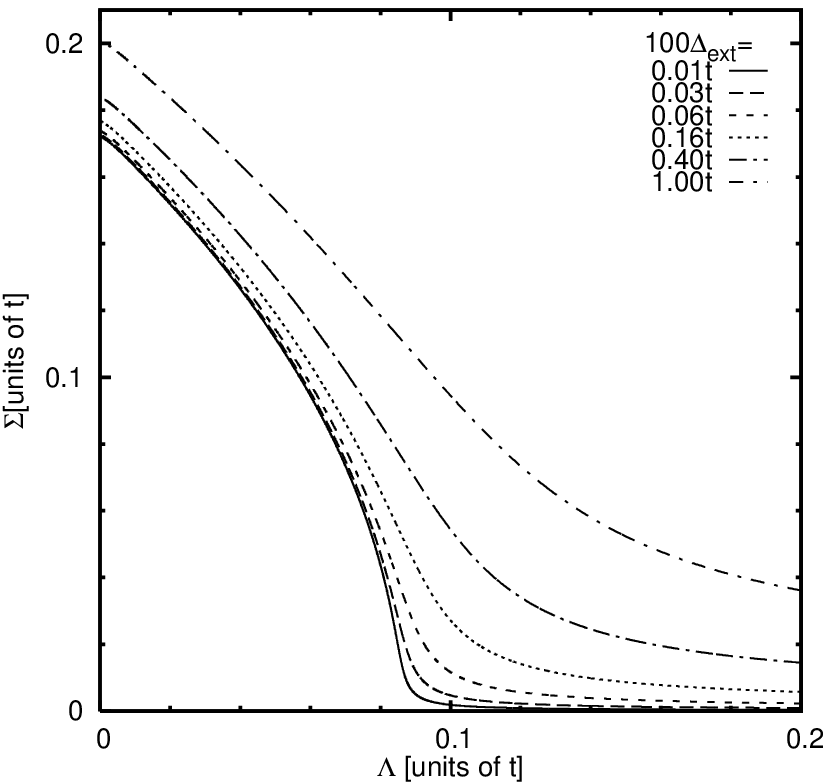}
  \includegraphics{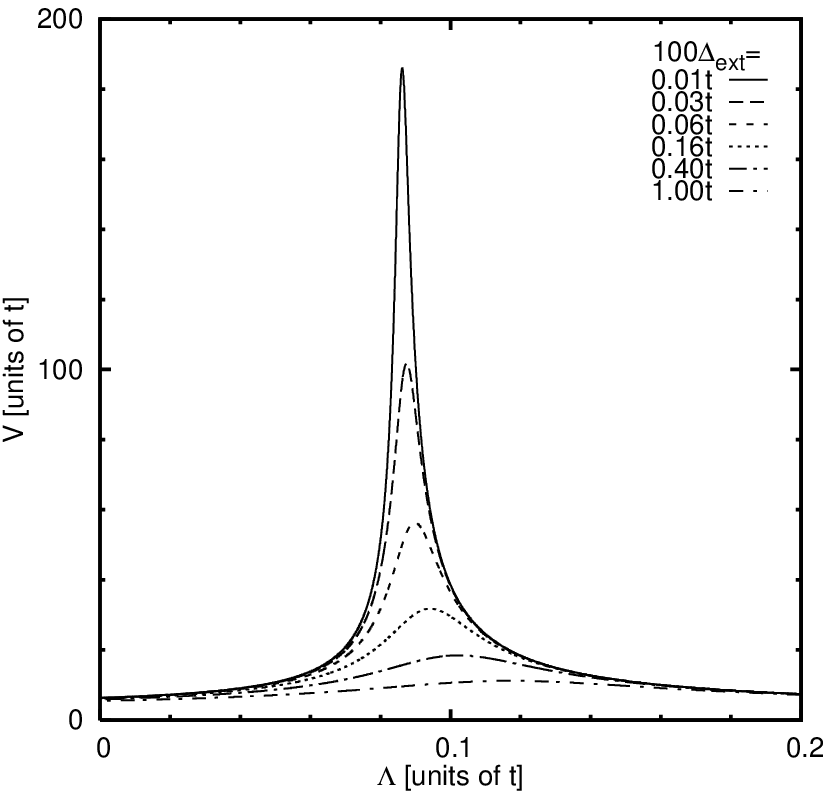}
  \caption{Low-energy portions of the $T=0$ 
    flows of the self-energy (top) and effective interaction (bottom) 
    for $V_0=2t$ and various $\Delta_{\rm ext}$. $\rho_0=1/2\pi$
    for all numerical calculations. Compare
    with Fig. \ref{fig:VRPA_T_U_0}.}
  \label{fig:flows_T0}
\end{figure}

We first consider zero temperature, $T=0$. 
In this case, it is easier to write down the analytical 
expressions, the roles of
the different right-hand side terms are easily identifiable, and the 
value of the effective interaction at $\Lambda=0$ is nearly 
independent of $\Delta_{\rm ext}$,
as $\Lambda_c$ is larger than the double width at half maximum of 
the effective interaction flow peak (see Fig. \ref{fig:flows_T0}, lower part).

In the limit $T\to 0$, \eqref{eq:RG_vertex_full} and
\eqref{eq:RG_sigma_full} become
\begin{align}
  \label{eq:RG_vertex_sharp}
  -\frac{\rm d}{{\rm d}\Lambda}V
  =&
  {V^2}
  \frac{\Lambda^2 \rho_0}{\sqrt{\Lambda^2+\Sigma^2}^3}
  -
  3V^2\Sigma\left(-\frac{{\rm d}\Sigma}{{\rm d}\Lambda}\right)
  \int_\Lambda^W{\rm d}\xi\frac{\xi^2\rho_0}{E^5},\\
  -\frac{\rm d}{{\rm d}\Lambda}\Sigma
  =&
  {V\Sigma}
  \rho_0
  \frac{1}{\sqrt{\Lambda^2+\Sigma^2}}.
  \label{eq:RG_sigma_sharp}
\end{align}
Notice that we have replaced the usual derivative with
respect to $s$ by $-\Lambda{\rm d}/{\rm d}\Lambda$. 
The typical shapes of this flow are exhibited in Fig.
\ref{fig:flows_T0}.
We see that the divergence of the effective interaction
is regularized by the initial symmetry-breaking field. 
Hence, we can integrate out all modes of the fermionic spectrum.
The four-point function in the CDW problem behaves analogously to the linear 
combination of normal and anomalous four-point functions in the BCS 
problem \cite{2004PThPh.112..943S} which drives the flow of the pairing 
amplitude.
The $\Lambda$-dependence of $\Sigma$ around the scale
where $V$ peaks approaches a kink for $\Delta_{\rm ext}\to 0$.
The graph of $\Sigma(\Lambda)$ resembles the graph
of $\Sigma(T)$, showing that temperature acts in a similar
fashion as the cutoff in this system. The main difference between 
the graphs is that $\Sigma(T\to 0)$ saturates exponentially while
$\Sigma(\Lambda\to 0)$ is linear.

The impact of the 
self-energy feedback on the RG flow can be thoroughly understood 
by analyzing the terms appearing in 
\eqref{eq:RG_vertex_sharp} and \eqref{eq:RG_sigma_sharp}.
Bear in mind that we think of the flow as progressing from
larger to smaller values of $\Lambda$, i.e. from right to
left in our plots. 
We first consider the limit $\Sigma=0$.
Now, the self-energy no longer flows while the effective interaction flows according to 
$-{\rm d}V/{\rm d}\Lambda\propto V^2/\Lambda$.
For any positive $V_0$, 
the solution of the corresponding initial
value problem is singular, showing that at $T=0$ symmetry is broken
for all repulsive initial interactions. 

An arbitrarily small initial symmetry-breaking field
immediately has dramatic consequences: As soon
as $V$ begins to increase strongly, so does the self-energy
due to the coupled effects of a large $V$ and a back-feeding
$\Sigma$ on the RHS of \eqref{eq:RG_sigma_sharp}. 
The second term on the RHS of \eqref{eq:RG_vertex_sharp} is negative. 
It corresponds to a correction of the flow by the modes which have been integrated over at $\Lambda' = \xi > \Lambda$ reflecting that their self-energy $\Sigma$ is changing in the flow.  
It becomes large as the slope of $\Sigma$ exceeds $\Sigma$ itself, while the positive first term
is damped by an effective $\Lambda^2/\Sigma^{3}$-dependence
as soon as $\Sigma$ becomes much larger than $\Lambda$.
The effective interaction is hence pulled
back down. It never reaches zero since
the negative second 
term on the RHS of \eqref{eq:RG_vertex_sharp} becomes
proportional to $V^3$ while the positive first term becomes
proportional to $V^2$ (${\rm d}\Sigma/{\rm d}\Lambda$ is $O(V)$). 
This implies that the self-energy never decreases. 
The two terms on the
RHS of \eqref{eq:RG_vertex_sharp} thus approach 
mutual cancellation, causing $V$ and in turn $-{\rm d}\Sigma/{\rm d}\Lambda$
to be almost constant in $\Lambda$ 
(the contributions from $\Sigma$
on the RHS of \eqref{eq:RG_sigma_sharp} cancel 
when $\Sigma \gg \Lambda$). 

The behavior described above remains qualitatively the same 
at finite temperatures,
as long as these are below the double width
at half maximum of $V^{\Delta_{\rm ext}}(T)$. This is illustrated by the 
lowest-temperature curve of the lower panel of Fig. \ref{fig:Lambda_crit_T_U_0}. The shape of $V(\Lambda)$
is very similar to the shape of $V(T)$ in the lower part of Fig.
\ref{fig:VRPA_T_U_0}. 
As it is found in many one-dimensional problems, temperature thus has an effect comparable to that of an energy or momentum cutoff, respectively, as implemented here. 

\section{Renormalization Group at Finite Temperatures}
\label{sec:RGfT}
We now turn to the analysis of the finite temperature fRG equations
\eqref{eq:RG_vertex_full} and \eqref{eq:RG_sigma_full}. Due to their
more involved nature, the analysis is largely numeric. 
We find the same behavior
as calculated using the conventional resummation methods applied in section 
\ref{sec:exact}. The initial symmetry-breaking field plays the same crucial
role as in the zero-temperature case.
We use a coupling of $V_0=2t$ in the following. 

\begin{figure}[htbp]
  \centering
  \includegraphics{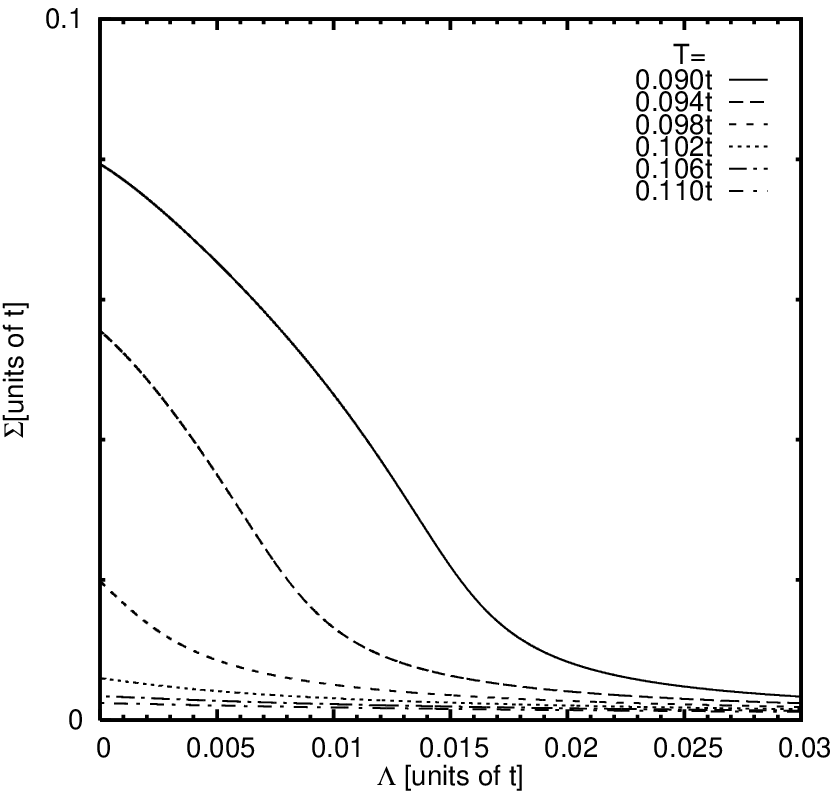}
  \includegraphics{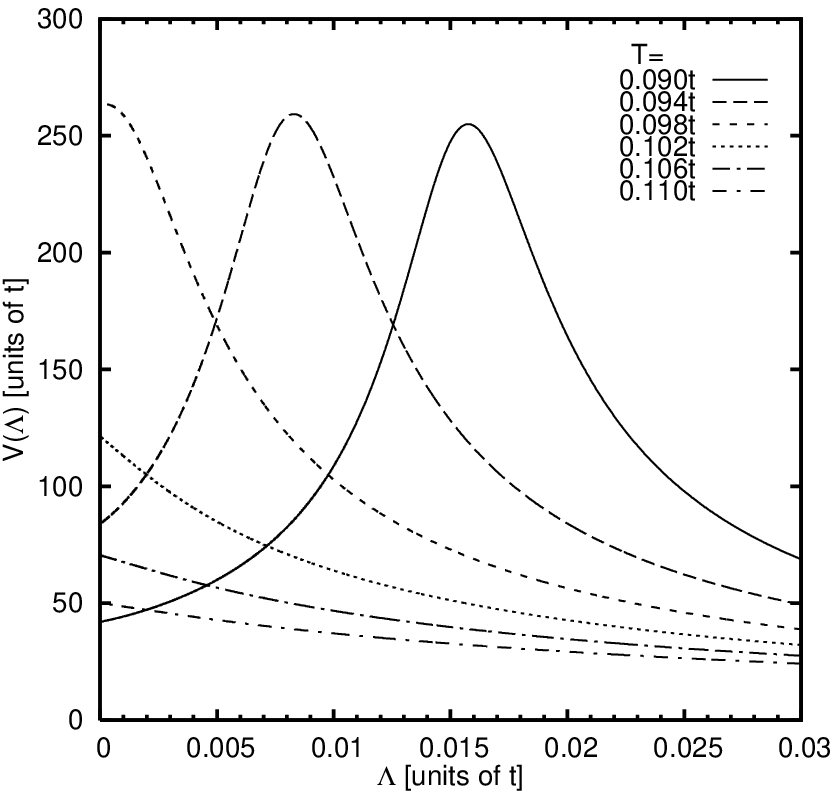}
  \caption{Flows of the self-energy $\Sigma$ and the effective interaction
    $V$ plotted against
    $\Lambda$ for temperatures around the transition, $V_0=2t$
    and $\Delta_{\rm ext}=10^{-4}t$. $\rho_0=1/2\pi$ for all
    numerical calculations.
  }
  \label{fig:Lambda_crit_T_U_0}
\end{figure}

The upper plot of Fig. 
\ref{fig:Lambda_crit_T_U_0} shows how the flow of the self energy flattens out when the temperature is increased beyond $T\approx 0.1t$. Above $T_c$, $\Sigma (\Lambda=0)$ vanishes in  the limit $\Delta_{\rm ext} \to 0$.
The RG flow of the effective interaction is shown in the lower panel of Fig.
\ref{fig:Lambda_crit_T_U_0}. The graph of the
flow is pushed to the left with increasing temperature, its shape
remaining largely unchanged. As the maximum approaches
zero, the final value of the effective interaction increases
until the maximum has reached $\Lambda=0$. This corresponds 
to the behavior of the effective interaction on the low-temperature side 
of Fig. \ref{fig:VRPA_T_U_0} 
(lower part). 
For even higher temperatures, the final value of the effective interaction
decreases, corresponding to the behavior on the high-energy side
of Fig. \ref{fig:VRPA_T_U_0}.

\begin{figure}[htbp]
  \centering
  \includegraphics{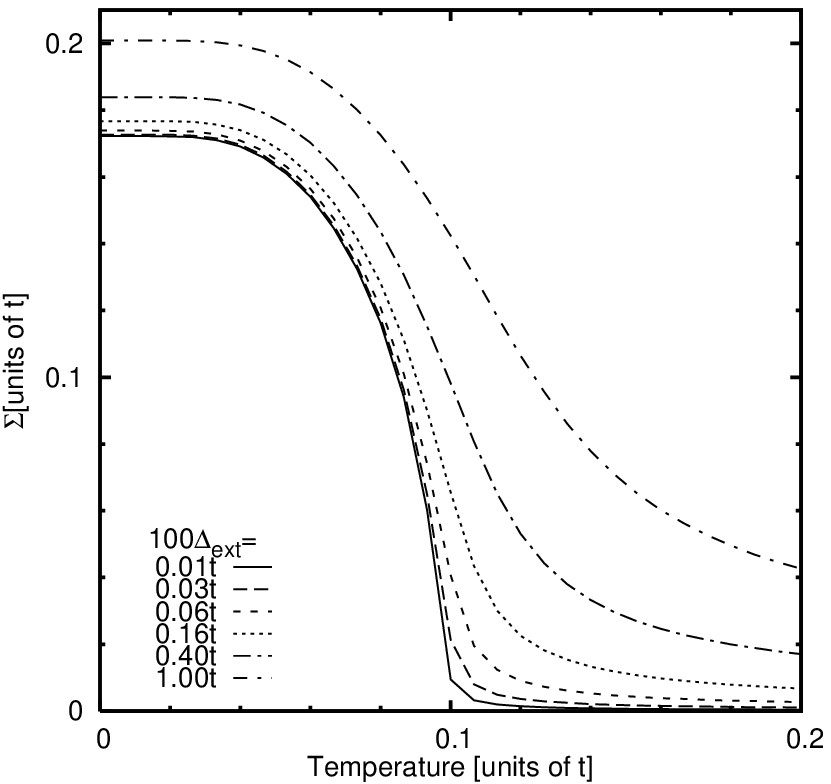}
  \includegraphics{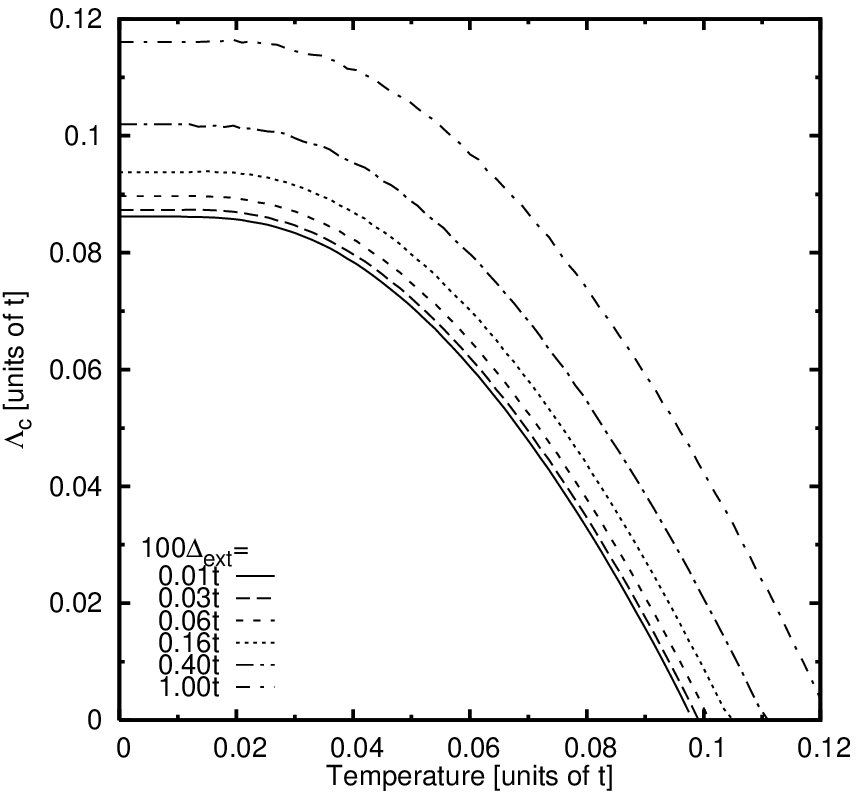}
  \caption{(top) Self-energy against temperature calculated
    using fRG methods for 
    $V_0=2t$ and small to intermediate $\Delta_{\rm ext}$.
    Minute deviations in comparison to Fig. \ref{fig:VRPA_T_U_0}
    are due to numerical errors in the integration
    of the fRG differential equations.
    (bottom) Position of the maximum of the cutoff-dependent 
    effective interaction
    against temperature for small to intermediate $\Delta_{\rm ext}$.
    $\rho_0=1/2\pi$ for all numerical calculations.}
  \label{fig:V_lambda_D_0}
\end{figure}

The lower part of Fig. \ref{fig:V_lambda_D_0} shows that
$\Lambda_c(T)$ saturates quickly
far below the critical temperature. It also illustrates the motion
of the effective interaction flow maximum, which is pushed to lower scales
by an increase in temperature as described above. 
It approaches zero in a linear
fashion with increasing temperature, in contrast to $\Sigma(T\to T_c)$,
which exhibits a square-root behavior.

We are furthermore interested in the dependence of the flow
of the self-energy at finite temperatures on $\Delta_{\rm ext}$.
In contrast to the resummation treatment case 
(see the upper part of Fig. \ref{fig:VRPA_T_U_0}), we cannot set
$\Delta_{\rm ext}$ to $0$ as the effective interaction would diverge before
the flow reaches zero scale. However, the upper part of Fig. \ref{fig:flows_T0}
implies that $\lim_{\Delta_{\rm ext}\to 0}\Sigma_{\Delta_{\rm ext}}(\Lambda)$
exists and is approached in a continuous fashion.  

\begin{figure}[htbp]
  \centering
  \includegraphics{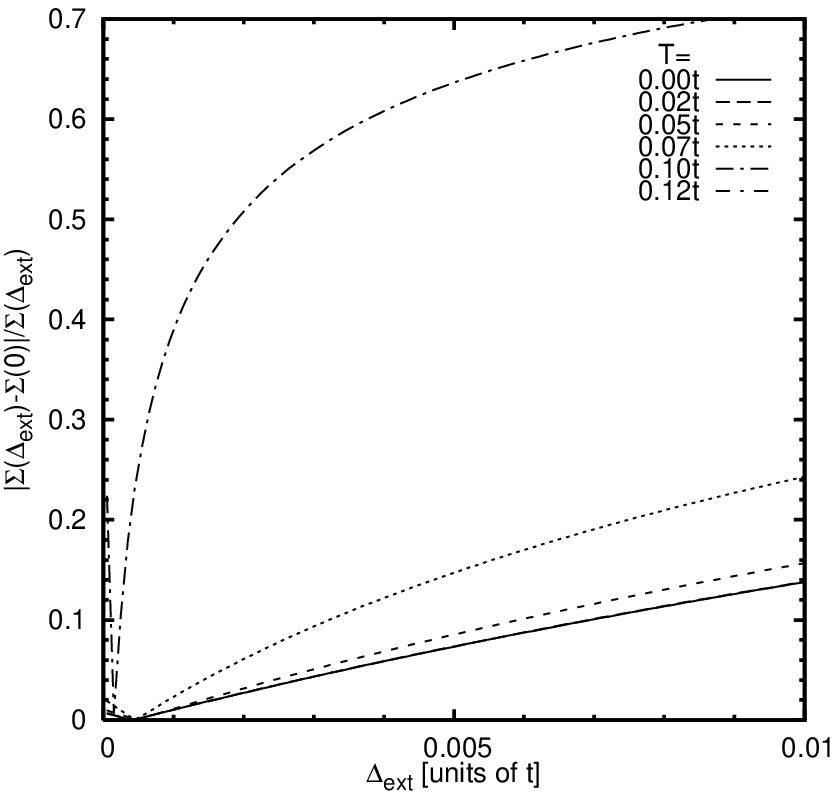}
  \includegraphics{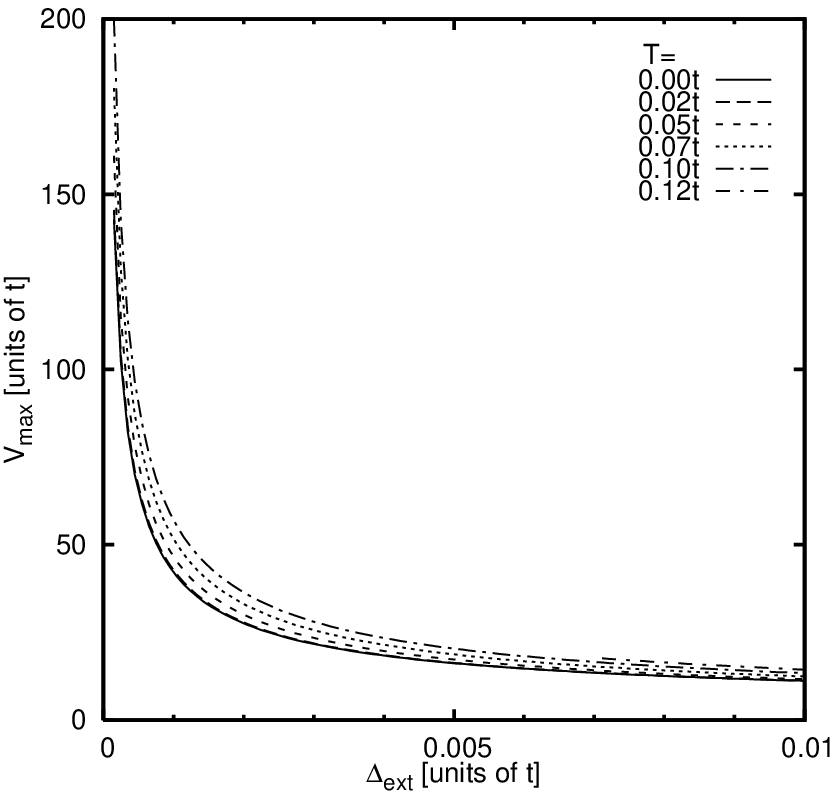}
  \caption{(top) Relative deviation of the fRG-derived self-energy from
    the value without external field against the initial anomalous 
    self-energy for temperatures from $T=0$ to just above $T_c$. 
    The values for $T=0.12t$ are $1$ and thus not visible in the selected
    range.
    (bottom) Maximal value of the effective interaction during the flow
    against the initial anomalous self-energy for temperatures
    from zero to just above $T_c$.
  }
  \label{fig:sigma_deviation}
\end{figure}

For $T>0$, if $\Delta_{\rm ext}$ is one or two percent of the final $\Sigma$, there is a clearly observable steep rise (see the upper part of Fig. \ref{fig:Lambda_crit_T_U_0}) which would allow a rather precise determination of $T_c$ even in more complicated models. 
However, note that for these values of $\Delta_{\rm ext}$, the effective interaction at the critical scale still reaches values $\sim 100t$ which are much larger than the bandwidth $4t$ (see, e.g., the lower part of Fig. \ref{fig:sigma_deviation}
for the temperature evolution). The agreement between fRG and mean-field results despite the fact that the effective interaction has grown to 
values large enough to render any perturbative
scheme useless in any general model 
underlines that the truncated fRG is exact for our model.

At the $T$-dependent critical
scale $\Lambda_c$, 
the effective interaction reaches a maximum whose height depends singularly
on $\Delta_{\rm ext}$ (see Fig. \ref{fig:sigma_deviation}). 
A numerical analysis shows that the maximal effective interaction is $\propto 1/\Delta_{\rm ext}^\alpha$ with $\alpha \approx 2/3$. 
The values for the maximal effective interaction can be read off for $T=0$ in 
Fig. \ref{fig:flows_T0} (bottom) and
for finite temperatures in the lower part of  Fig. \ref{fig:sigma_deviation}.
For a straightforward application of the method to  non-mean-field models, 
it may be necessary to restrict the maximum of the effective interaction to smaller 
values to justify a perturbative treatment.
Then the rise of $\Sigma_{\Lambda=0} (T)$ at $T_c$ gets 
smeared out strongly
(see Fig. \ref{fig:V_lambda_D_0} (upper part) and Fig. \ref{fig:VRPA_T_U_0} 
(upper part)). 
The upper part of Fig. \ref{fig:sigma_deviation} shows
that far below $T_c$, the relative error $\Delta\Sigma/\Sigma$
is linear in $\Delta_{\rm ext}$. At the highest considered
$\Delta_{\rm ext}$, it is already of the order of $0.1$ while
$V(\Lambda_c)$ is still larger than two times the bandwidth.
     
\section{Conclusion}
\label{sec:conclusion}
We have applied the functional renormalization group to a simple model for the CDW transition of spinless fermions on a hypercubic lattice in $d$ spatial dimensions. The main purpose was to test and further understand the recently developed method for the fRG flow into symmetry-broken phases in the case of a broken discrete symmetry and to generalize it to finite temperatures. 

Earlier RG treatments of low-dimensional interacting fermions generically encountered flows to strong coupling where a class of coupling constants becomes 
too large to be treatable perturbatively. 
Then, the flow had to be stopped. While these flows still contain 
ample information about the low-energy physics of the system, a continuation down to lowest scales is desirable to obtain a controlled access to possible symmetry-broken phases. A recent treatment of the BCS model showed that the divergence of the interactions at finite RG scale can be avoided by introducing a small initial symmetry-breaking field. Furthermore, a correct treatment of the flow of the corresponding anomalous self-energy yields the correct value for the energy gap at zero scale. The interaction vertex corresponding to the Goldstone boson in the case of breaking of the continuous U(1) symmetry still becomes large $\propto 1/\Delta_{\rm ext}$ when the initial symmetry-breaking field $\Delta_{\rm ext}$ is sent to zero. 

Here we applied this method to the CDW problem with a half-filled band. Including a small initial anomalous CDW self-energy $\Delta_{\rm ext}\not= 0$ we found that the flow of the effective interaction goes through a maximum at the critical scale $\Lambda_c$ and falls off below. 
In this case, the effective interaction does not become large as $\Lambda \to 0$ and no anomalous interactions are generated, reflecting the fact that the discrete symmetry breaking of the commensurate CDW does not produce a Goldstone mode.  
% RG 03.11.05
%At $\Delta_{\rm ext}$, the anomalous self-energy $\Sigma$ corresponding 
At $\Lambda_c$, the anomalous self-energy $\Sigma$ corresponding 
to the strength of the discrete symmetry breaking rises steeply from its initial value $\Delta_{\rm ext}$ and finally approaches a finite value for $\Lambda \to 0$. 
%RG 03.11.05
% Letting $\Delta_{\rm ext}\to 0 $ 
% we found that $\Sigma_{\Lambda=0}$ converges
% to the exact mean-field result for the case without
%  explicit symmetry breaking $\Delta_{\rm ext}=0$.
Letting $\Delta_{\rm ext}\to 0 $, 
we found that $\Sigma_{\Lambda=0}$ converges to the exact result
from the case without explicit symmetry breaking, $\Delta_{\rm ext}=0$.

Studying the flow at nonzero temperatures, we have
noted that the temperature dependence of the critical scale differs qualitatively from the temperature dependence of the offdiagonal self-energy, i.e. no square-root like rise of $\Lambda_c (T)$ is found below $T_c$. 
Therefore, $\Lambda_c (T)$ is only a rough estimate for the self-energy.
Thus, a proper treatment of the flow of the self-energy into 
the symmetry-broken phase is needed in order to assess the temperature
dependence of the spectral gap.

Similar to the BCS problem, the maximal value of the effective 
interaction at $\Lambda_c$ depends crucially on the explicit 
symmetry breaking $\Delta_{\rm ext}$.
%RG 03.11.05
For the mean-field model we have investigated in the present work,
the truncated RG scheme is exact. Thus, large effective
interactions do not cause problems.
However, for the practical application of the fRG scheme to non-mean-field 
%For the practical application of the fRG scheme to non-mean-field 
%
models where our truncated RG scheme is not exact it is necessary 
that the interactions do not become too large at any scale. 
Otherwise the feedback of the interactions in the ordering channel on processes in other channels would become large and the perturbative RG would break down. 
In the case of  breaking of a discrete symmetry studied here, the only region where the interactions become large at zero temperature is near $\Lambda \approx \Lambda_c$. There, however, our explicit calculation shows that if we want to stay close to the mean-field results without explicit symmetry breaking we cannot choose too large a $\Delta_{\rm ext}$. This in turn leads to maximal values of the effective interaction which are much larger than the bandwidth. If one wants to work  with couplings which stay within the order of the bandwidth (or expressed differently, with dimensionless coupling less than one), the deviations from the $\Delta_{\rm ext}=0$ results are rather large.   At finite temperatures, the explicit symmetry breaking smears out the transition. Again, the onset of the growth of the off-diagonal self-energy only allows for a reasonable definition of a critical temperature when the maximal effective interaction is allowed to get large. 

%RG 03.11.05
%On first sight these results appear to represent severe limitations 
%for the applicability of our method to non-mean-field models. 
On the one hand, the large effective interactions discussed above
could spoil the application of this truncated RG scheme to 
non-mean-field models.
%On the other hand, one should keep in mind that large effective
%interactions only occur for a narrow range of frequencies and
On the other hand, one should keep in mind that in non-mean-field
models large effective
interactions only occur for a narrow range of frequencies and
momenta, if the bare interaction is weak. Hence, the feedback
of these large interactions into other interaction channels
will be strongly reduced by phase-space factors.
It will be interesting to see how this physics saves or destroys 
for more general types of interactions the mean-field 
picture of the reduced models.
In particular, the space dimension plays a crucial role in this respect. Understanding these issues could provide insights into the emergence of short-range ordered states in a weak coupling picture.

With the steps outlined in this article, we believe that the fRG flow into symmetry-broken states is well-un\-der\-stood for mean-field-type models. In the future it can be generalized to more complicated models, where  it will reveal its full usefulness. 
Finally, we note that an alternative method for the flow into the
symmetry-broken phase based on a two-particle irreducible RG formalism has
been put forward by Dupuis \cite{dupuis}. It will be interesting to explore the
applicability of these new RG schemes in more realistic models.
 
We thank Tilman Enss,  Andrey Katanin, Julius Reiss
and Manfred Salmhofer for useful
discussions.

\end{document}